# Mechanical Characterization of Brain Tissue in Compression at Dynamic Strain Rates

Badar Rashid[a], Michel Destrade[b,a], Michael Gilchrist[a*]

[a]School of Mechanical and Materials Engineering, University College Dublin, Belfield, Dublin 4, Ireland

[b]School of Mathematics, Statistics and Applied Mathematics, National University of Ireland Galway, Galway, Ireland

*Corresponding Author

Tel: + 353 1 716 1884/1991, + 353 91 49 2344  Fax: + 353 1 283 0534

Email: Badar.Rashid@ucdconnect.ie (B. Rashid), michael.gilchrist@ucd.ie (M.D. Gilchrist), michel.destrade@nuigalway.ie (M. Destrade)

**Abstract**   Traumatic brain injury (TBI) occurs when local mechanical load exceeds certain tolerance levels for brain tissue. Extensive research has been done previously for brain matter experiencing compression at quasistatic loading; however, limited data is available to model TBI under dynamic impact conditions. In this research, an experimental setup was developed to perform unconfined compression tests and stress relaxation tests at strain rates ≤ 90/s. The brain tissue showed a stiffer response with increasing strain rates, showing that hyperelastic models are not adequate. Specifically, the compressive nominal stress at 30% strain was 8.83 ± 1.94, 12.8 ± 3.10 and 16.0 ± 1.41 kPa (mean ± SD) at strain rates of 30, 60 and 90/s, respectively. Relaxation tests were also conducted at 10% - 50% strain with the average rise time of 10 ms, which can be used to derive time dependent parameters. Numerical simulations were performed using one-term Ogden model with initial shear modulus $\mu_o$ = 6.06 ± 1.44, 9.44 ± 2.427 and 12.64 ± 1.227 kPa (mean ± SD) at strain rates of 30, 60 and 90/s, respectively. A separate set of bonded and lubricated tests were also performed under the same test conditions to estimate the friction coefficient $\mu$, by adopting combined experimental – computational approach. The values of $\mu$ were 0.1 ± 0.03 and 0.15 ± 0.07 (mean ± SD) at 30 and 90/s strain rates, respectively, indicating that pure slip conditions cannot be achieved in unconfined compression tests even under fully lubricated test conditions. The material parameters obtained in this study will help to develop biofidelic human brain finite element models, which can subsequently be used to predict brain injuries under impact conditions.

# 1 Introduction

The human head has been identified as the most sensitive region frequently involved in life-threatening injuries such as road traffic accidents, sports accidents and falls. Intracranial brain deformation caused by rapid angular acceleration or blunt impact to the head during injurious events is responsible for traumatic brain injuries (TBIs), which are a leading cause of death or disability. TBI involves acute subdural hematoma, brain contusion and diffuse axonal injury. To gain a better understanding of the mechanisms of TBI, several research groups have developed numerical models which contain detailed geometric descriptions of anatomical features of the human head, in order to investigate internal dynamic responses to multiple loading conditions (Ho and Kleiven, 2009; Horgan and Gilchrist, 2003; Kleiven, 2007; Kleiven and Hardy, 2002; Ruan et al., 1994; Takhounts et al., 2003; Zhang et al., 2001). However, the fidelity of these models is highly dependent on the accuracy of the material properties used to model biological tissues; therefore, more systematic research on the constitutive behavior of brain tissue under impact is essential.

Over the past three decades, several research groups investigated the mechanical properties of brain tissue in order to establish constitutive relationships over a wide range of loading conditions. Recently, Chatelin et al. (2010) carried out a comprehensive review on fifty years of brain tissue mechanical testing and compared *in vitro* and *in vivo* results in order to analyze the difficulties linked with *in vitro* experimental protocols and the advantages of using recently developed non-invasive *in vivo* techniques. Mostly dynamic oscillatory shear tests were conducted over a frequency range of 0.1 to 10000 Hz (Arbogast et al., 1997; Bilston et al., 2001; Brands et al., 2004; Darvish and Crandall, 2001; Fallenstein et al., 1969; Hrapko et al., 2006; Nicolle et al., 2004; Nicolle et al., 2005; Prange and Margulies, 2002; Shuck and Advani, 1972; Thibault and Margulies, 1998) and unconfined compression tests (Cheng and Bilston, 2007; Estes and McElhaney, 1970; Miller and Chinzei, 1997; Pervin and Chen, 2009; Prange and Margulies, 2002; Tamura et al., 2007), while a limited number of tensile tests (Miller and Chinzei, 2002; Tamura et al., 2008; Velardi et al., 2006) were performed and the reported properties vary from study to study.

Brain tissue consists of gray and white matter and is covered with the thin layer of pia and arachnoid membranes. The gray matter is made up of neuronal cell bodies, which are distributed at the surface of the cerebral cortex which does not seem to have any directional preference, while the white matter is composed of bundles of myelinated nerve cell processes (or axons), that can be highly oriented (Nicolle et al., 2004). The white matter can be considered as a transversely isotropic structure whereas the gray matter is simply isotropic (Arbogast and Margulies, 1999; Nicolle et al., 2004). The recorded variation in test results is probably related to the anisotropic and inhomogeneous nature of brain tissue and the broad range of test conditions. Depending on the application, viscoelastic and even purely elastic models have been used in different analyses by various research groups. The characteristic time scale is very important for choosing the material model. Impact usually is modeled with viscoelasticity, while long term processes like hydrocephalus can be modeled using poroelasticity or mixture theory due to the need to account for interstitial fluid movement (Kyriacou et al., 2002). In fact, the duration of a typical head impact is of the order of milliseconds. Therefore to model TBI, we need to characterize brain tissue properties over the expected range of loading rate appropriate for potentially injurious circumstances.

In the case of unconfined compression of brain tissue at intermediate velocities, limited studies have been conducted (Estes and McElhaney, 1970; Pervin and Chen, 2009; Tamura et al., 2007). Estes and McElhaney (1970) performed in-vitro unconfined compression tests on human and rhesus monkey brain tissues up to 50% engineering strain at strain rates of 0.08, 0.80, 8.0 and 40 /s. It is the strain rate of 40/s that is closest to impact conditions. Similarly, Pervin and Chen (2009) performed tests at strain rates of 0.01, 0.1, 1000, 2000, 3000/s, where the mechanical properties determined at high strain rates (1000 – 3000/s) are associated with blast loading incidents, such as penetrating gunshot injuries to the head and open skull blast brain injuries. Miller (1999) and Miller and Chinzei (1997) performed unconfined compression tests on brain tissue, but at slow loading velocities (0.005, 5, 500 mm/min) and proposed linear and nonlinear viscoelastic models to describe deformation behavior under compression, suitable for neurosurgical procedures. Similarly, Tamura et al. (2007) conducted a study at strain rates of 1, 10, 50/s, where the fastest rate was closest to impact speeds. Of these various results it is those at strain rates in the range of 10 – 100/s and at compressive strain levels of 10 – 50% that are of direct relevance to impact injury.

Extensive research is still underway to understand the biomechanics of TBI. Diffuse axonal injury (DAI) is a type of TBI which is characterized by microscopic damage to axons throughout the white matter of the brain, as well as focal lesions in the corpus callosum and rostral brainstem. DAI in animals and human has been hypothesized to occur at macroscopic shear strains of 10% – 50% and strain rates of approximately 10 – 50/s (Margulies et al., 1990; Meaney and Thibault, 1990) although locally the strains and strain rates could be much higher than their macroscopic values due to the complex geometry and material inhomogeneities of brain tissue. Several studies have been conducted to determine the range of strain and strain rates associated to DAI. Bain and Meaney (2000) investigated *in vivo*, tissue-level*,* mechanical thresholds for axonal injury by developing a correlation between the strains experienced in the guinea pig optic nerve and the morphological and functional injury. The threshold strains predicted for injury ranged from 0.13 to 0.34. Similarly, Pfister et al. (2003) developed a uniaxial stretching device to study axonal injury and neural cell death by applying strains within the range of 20%–70% and strain rates within the range of 20 – 90/s to create mild to severe axonal injuries. Bayly et al. (2006) carried out *in vivo* rapid indentation of rat brain to determine strain fields using harmonic phase analysis and tagged MR images. Values of maximum principal strains > 0.20 and strain rates > 40/s were observed in several animals exposed to 2mm impacts of 21 ms duration. Studies conducted by Morrison et al. (2000, 2003, 2006), also suggested that the brain cells are significantly damaged at strains > 0.10 and strain rates > 10/s.

Considering the difficulty of obtaining human brain tissue for *in vitro* testing, experiments are usually performed on animal brain samples (monkey, porcine, bovine, rabbit, calf, rat or mouse). Galford and McElhaney (1970) showed that shear, storage and loss moduli are 1.5, 1.4 and 2 times higher for monkeys than for humans, respectively. Similarly, Estes and McElhaney (1970) performed in-vitro unconfined compression tests on human and Rhesus monkey tissue and found that the response of the Rhesus monkey tissue was slightly higher than the response of human brain tissue at comparable compression rates. Differences between human and porcine brain properties were also pointed out by Prange et al. (2000) who demonstrated that human brain tissue stiffness was 1.3 times higher than that of porcine brain. However, Nicolle et al. (2004) observed no significant difference between the mechanical properties of human and porcine brain matter. Pervin and Chen (2011) found no difference between the *in vitro* dynamic mechanical response of brain matter in different animals (porcine, bovine and caprine), different breeds and different genders. Based on the similar behavior of brain tissue in different species as observed by Nicolle et al. (2004) and Pervin and Chen (2011), the properties of porcine brain tissue may be used in the human finite element head models.

The present study has been carried out to determine mechanical properties of brain tissue by performing compression tests at 30, 60 and 90/s strain rates up to 30% strain. Relaxation tests were performed at 10%, 20%, 30%, 40%, 50% strain with an average rise time of 10 ms to determine time dependent material parameters. The challenge with these tests was to attain uniform velocity during the compression phase of the brain tissue. This was achieved by developing a test apparatus that provided uniform velocity during unconfined compression of brain tissue. The experimental data was used to estimate Ogden, Fung and Gent material parameters. Next, numerical simulations were performed using ABAQUS 6.9/Explicit after applying boundary conditions to mimic experimental test conditions. A separate set of bonded and lubricated tests were also performed under the same test conditions to estimate friction coefficient, $\mu$ by adopting combined experimental – computational approach. This study provides new insight into the behavior of brain tissue under dynamic impact conditions, which will assist in developing effective brain injury criteria and adopting efficient countermeasures against TBI.

## 2  Materials and Methods

### 2.1  Specimen Preparation

Fifteen fresh porcine brains were collected from a local slaughter house and tested within 8 h, which was consistent with previous work (Estes and McElhaney, 1970; Miller and Chinzei, 1997). Each brain was preserved in a physiological saline solution (0.9% NaCl /154 mmol/L) at 4 to 5$^{o}$C during transportation. All samples were prepared and tested at a room temperature ~ 22 $^{o}$C and relative humidity of 55 – 65%. The dura and arachnoid were removed and the cerebral hemispheres were first split into right and left halves by cutting through the corpus callosum.

As shown in Fig. 1, one half of the cerebral hemisphere was cut in the coronal plane and cylindrical specimens were extracted while cutting from an anterior – posterior direction. Similarly, further slices in the coronal plane were also made from each cerebral hemisphere.

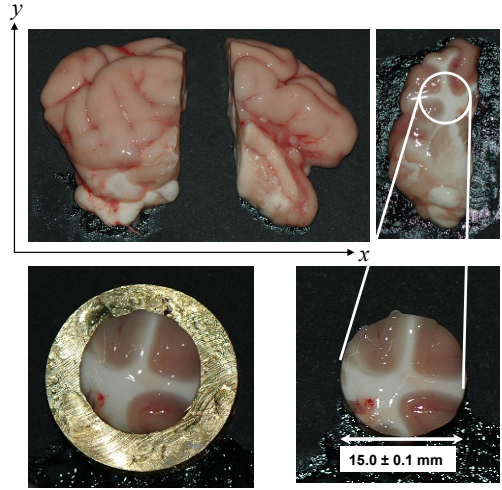

Fig. 1 – Cylindrical brain specimen (15.0 ± 0.1 mm diameter) excised from the anterior – posterior direction, perpendicular to coronal plane.

In Fig.1, the specimens excised from the brain were composed of mixed white and gray matter. Cylindrical samples of nominal diameter 15.0±0.1 mm were cut using a circular steel die cutter of 15.5 mm diameter and then inserted in a cylindrical metal disk with 15.2 mm internal diameter and 5.1 mm thickness as shown in Fig.1. The excessive brain portion was then removed with a surgical scalpel to maintain an approximate specimen height of 5.0±0.1 mm. A visible contraction of the cylindrical samples occurred immediately after they were removed from the brains, revealing the presence of residual stresses *in-vivo*. When measuring the dimensions of the specimens, it was noted that the nominal dimensions were reached after a few minutes; it was at this stage that testing commenced. The actual diameter and height of specimens measured before the testing were 15.1±0.1 mm and 5.0±0.2 mm (mean ± SD). 28 specimens from 7 brains for the unconfined compression tests and 64 specimens from 8 brains for the stress relaxation tests were prepared. The time elapsed between harvesting of the first and the last specimens from each brain was 16 ~ 20 minutes for the unconfined compression tests, however, 35 ~ 45 minutes elapsed for the stress relaxation tests. Physiological saline solution was applied to specimens frequently during cutting and before the tests in order to prevent dehydration. The specimens were not all excised simultaneously, rather each specimen was tested first and then another specimen was extracted from the cerebral hemisphere. This procedure was important to prevent the tissue from losing some of its stiffness and to prevent dehydration, and thus contributed towards repeatability in the experimentation.

## 2.2    Experimental Setup

Unconfined compression tests were performed on cylindrical brain specimens by using the test apparatus shown in Fig. 2. To achieve uniform velocity during the compression of tissue in the axial direction, the test apparatus was purposely developed and calibrated at high velocities. A programmable servo motor with a controlled electronic actuator (LEY 16 A) was used. This has a maximum speed of 500 mm/s and a 50 mm stroke with a positioning repeatability of ± 0.02 mm. The specimen stage was raised to the point where the compression platen completes approximately 50% of its stroke length.

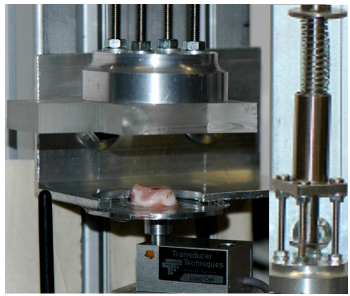

(a)

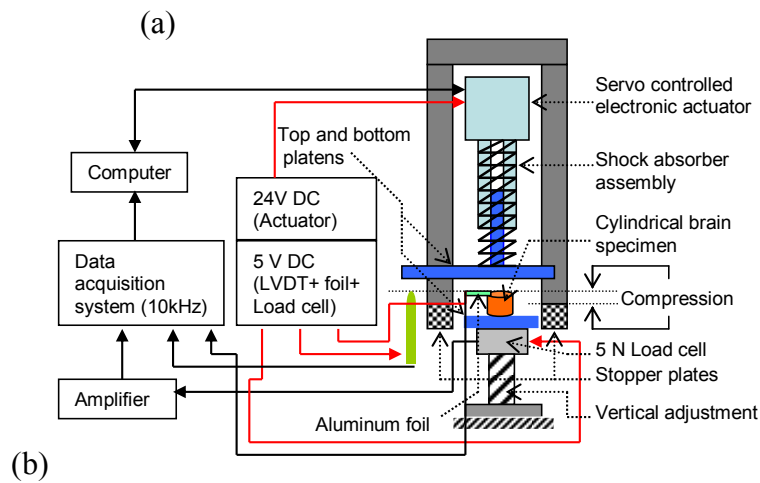

(b)

Fig. 2 – Experimental setup for the unconfined compression of brain tissue (a) cylindrical brain specimen before compression, (b) schematic diagram of complete test apparatus. Red lines indicate inputs to electrical assemblies and black lines indicate data output to computer.

To avoid any damage to the electronic assembly during the experiments, a shock absorber assembly with a compression platen and stopper plate was developed, as shown in Fig. 2. The stopper plate was fixed on the machine column to stop the compression platen at the middle of the stroke with an impact during the uniform velocity phase, while the actuator completes its travel without producing any backward thrust to the servo motor and other components. A GSO series -5 to +5 N load cell (Transducer Techniques) was used for the measurement of compressive force. This had a rated output of 1 mV/V nominal and safe overload of 150% of rated output. The load cell was calibrated against known masses and a multiplication factor of 13.62 N/V (determined through calibration) was used for the conversion of voltage to load. An integrated single-supply instrumentation amplifier (AD 623 G =100, Analog Devices) with built-in single pole low-pass filter having cut-off frequency of 10 kHz was used. The out put of the amplifier was passed through second single pole low-pass filter with cut-off frequency of 16 kHz. The amplified signal was analyzed through a data acquisition system with a sampling frequency of 10 kHz. Because of the probable viscoelastic nature of brain tissue, it was necessary to measure displacement of the compression platen with high precision. The linear variable displacement transducer (LVDT) was used to measure displacement during the unconfined compression phase. The type – ACT1000A LVDT developed by RDP electronics had a sensitivity; 16 mV/mm (obtained through calibration), range ± 25 mm, linearity ± 0.25 percent of full range output, spring force at zero position 2.0 N and spring rate of 0.3 N/cm.

## 2.3    Synchronization of Force and Displacement Signals

In this experimental setup, the displacement sensor is not in contact with the platen throughout the test, however the tip of the sensor compresses upon collision. In order to measure displacement with precision, a cylindrical specimen of nominal dimensions (15.0 mm diameter and 5.0 mm thick) was prepared from commercially available silicone-based gel Sylgard-527 (Dow Corning Corporation, Midland, MI, USA) for the calibration of the setup. Moreover, in order to capture the instant of contact between the platen and the specimen, a 5V DC circuit including 10 kΩ resistance was developed. The +ve terminal was attached with aluminum foil (to be placed on the Sylgard gel specimen) and −ve terminal was grounded with the compression platen. A voltage drop signal was initiated when the top platen established contact with the specimen during the compression phase. Outputs from both terminals were also introduced into the data acquisition system.

After a small number of trials, the displacement (mm) signal from the LVDT, the force (N) signal from the load cell and the voltage drop signal from the aluminum foil were directly captured through the data acquisition system. This made it possible to precisely locate the position of the displacement sensor and to identify the zero strain point as shown in Fig.3. It was also possible to find the equilibrium position of the sensor as indicated by point B (Fig. 3) to measure the total displacement of the sensor. At this preparatory stage, calibration was done without brain tissue to record the displacement – time signal to ensure uniform velocity at strain rates of 30, 60 and 90/s.

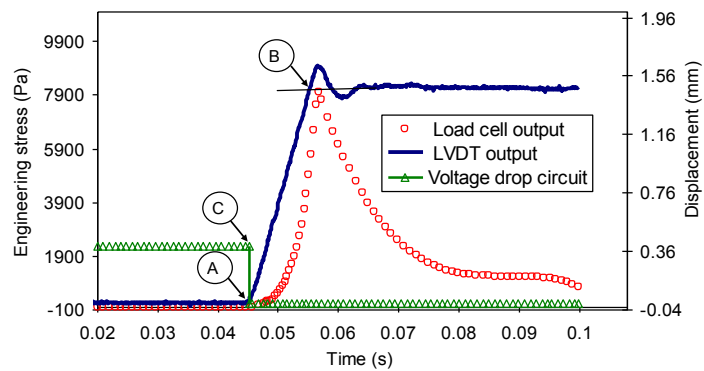

Fig. 3 – A typical output from the data acquisition system indicating three signals. Point A indicates synchronization and initiation of all three signals, point B is the equilibrium position of the displacement sensor and point C indicates voltage drop when the top platen establish contact with the specimen.

Experimentation with the brain tissue specimen started after the setup was fully calibrated at a particular velocity. Ideally, the force and displacement signal should initiate with zero time difference, which was very difficult to attain because of the brain specimen thickness variations (5.0 ± 0.1 mm). The change in diameter in the middle of the specimen during the compression of the specimen was measured with the help of recorded images obtained from a high speed digital camera (Phantom V5.1, CMOS 10 bit Sensor, 1200 frames per second (fps) at maximum resolution - 1024 x 1024 and 95000 fps at a minimum resolution - 64 x 32). In the present study, the high speed image recording of brain tissue during compression tests was done at a frame rate of 3906 fps with 640 x 480 resolutions.

## 2.4 Compression Tests

Unconfined compression tests were performed on mixed white and gray matter on cylindrical specimens up to 30% strain. 28 specimens were extracted from 7 brains (4 specimens from each brain), from the coronal plane excised in the anterior – posterior direction. The velocity of the compression platen (top platen) was adjusted to 150, 300 and 450 mm/s corresponding to approximate strain rates of 30, 60 and 90/s, respectively. The attainment of uniform velocity was also confirmed during the calibration process. The top and lower platens were thoroughly lubricated with Phosphate Buffer Saline (PBS 0.9% NaCl /154 mmol/L) solution, before every test. The solution was used to minimize frictional effects and to ensure, as much as possible, uniform expansion in the radial direction. Each specimen was tested once and then discarded because of the highly dissipative nature of brain tissue. No preconditioning was performed due to the extreme delicacy and adhesiveness of brain tissue. All tests were conducted at a room temperature ~ 22 °C and relative humidity of 55 – 65%.

## 2.5 Force Relaxation Tests

A separate set of force relaxation experiments were performed on cylindrical specimens of mixed white and gray matter (4.0 ± 0.1 mm thick and 15.0 ± 0.1 mm diameter). Here, 64 specimens were extracted from 8 brains (4 samples from each cerebral hemisphere), from the coronal plane excised in the anterior – posterior direction. Ten force relaxation tests were performed at each 10%, 20%, 30%, 40% and 50% strain in order to investigate the behavior of brain tissue to a step-like strain at variable strain magnitudes. The specimens were compressed from 120 – 500 mm/s to various strain levels with a sampling rate of 10 kHz. The average rise time measured from the force relaxation experiments was 10 milliseconds (ms). Force vs. time data was recorded up to 500 ms. Force relaxation experiments at intermediate strain rates are very important for the determination of time dependent parameters such as $\tau_k$, the characteristic relaxation times, and $g_k$, the relaxation coefficients. These are required to simulate impact conditions related to TBI. These parameters can then be used directly in a suitable constitutive model for the determination of stress, taking into account the strain rate dependency of the material.

## 2.6  Bonded and Lubricated Compression Tests

A separate set of bonded and lubricated experiments were performed on brain tissue in order to estimate the amount of friction generated during unconfined compression tests. 24 specimens were excised from 12 porcine brains (2 specimens from each brain) according to the procedure discussed in Sections 2.1 and 2.4. 6 bonded and 6 lubricated tests were performed at each strain rate of 30/s and 90/s. Since two specimens were extracted from the same brain, one specimen was utilized for the lubricated unconfined compression tests while the other was used for the bonded unconfined compression tests. In the bonded tests, the surfaces of the platens were first covered with a masking tape substrate to which a thin layer of surgical glue (Cyanoacrylate, Low-viscosity Z105880–1EA, Sigma-Aldrich, Wicklow, Ireland) was applied. The prepared cylindrical specimen of brain tissue was placed on the lower platen. After approximately one minute settling time, the top platen was allowed to travel at the required velocity. A thin layer of surgical glue applied on the platen surfaces fully restricted lateral expansion of the top and bottom surfaces of the brain specimen. This procedure was necessary to minimize the possibility of errors between the lubricated and bonded experiments in a controlled environment. Detailed analysis on bonded compression is discussed in a separate study.

# 3  Phenomenological Constitutive Models

## 3.1  Preliminaries

In general, an isotropic hyperelastic incompressible material is characterized by a strain-energy density function $W$ which is a function of two principal strain invariants only: $W = W(I_1, I_2)$, where $I_1$ and $I_2$ are defined as (Ogden, 1997)

$$I_1 = \lambda_1^2 + \lambda_2^2 + \lambda_3^2, \tag{1}$$

$$I_2 = \lambda_1^2 \lambda_2^2 + \lambda_1^2 \lambda_3^2 + \lambda_2^2 \lambda_3^2. \tag{2}$$

Here $\lambda_1^2, \lambda_2^2, \lambda_3^2$ are the squares of the principal stretch ratios, linked by the relationship $\lambda_1 \lambda_2 \lambda_3 = 1$, due to incompressibility.

As discussed previously, an effort was made to ensure that the samples under compression expand uniformly; therefore, it was first assumed that under unconfined compression the deformation was homogenous. Then the Eulerian and Lagrangian principal axes of strain and stress are aligned with the direction of compression, $x_1$, say, and with any two orthogonal axes (lateral) $x_2, x_3$, say. Due to symmetry and incompressibility, the stretch ratios are now of the form

$$\lambda_1 = \lambda, \quad \lambda_2 = \lambda_3 = \frac{1}{\sqrt{\lambda}} \tag{3}$$

where $\lambda \leq 1$ is the stretch ratio in the direction of compression. Also, Eqs.(1) and (2) give

$$I_1 = \lambda^2 + 2\lambda^{-1}, \quad I_2 = \lambda^{-2} + 2\lambda, \tag{4}$$

so that $W$ is now a function of $\lambda$ only. During the experimental compression tests, the principal stretch ratio $\lambda$ was calculated from the measure of the elongation $e$ using equation: $\lambda = 1 - e$. The nominal/Lagrange stress component along the direction of compression $S_{11}$ was evaluated as $S_{11} \equiv F/A$, where $F$ is the compressive force, as measured in Newtons by the load cell, and $A$ is the area of a cross section of the sample in its undeformed state. The experimentally measured nominal stress was then compared to the predictions of the hyperelastic models from the relation (Ogden, 1997),

$$S_{11} = \frac{d\widetilde{W}}{d\lambda}, \text{ where } \widetilde{W}(\lambda) \equiv W(\lambda^2 + 2\lambda^{-1}, \lambda^{-2} + 2\lambda), \tag{5}$$

and the material parameters were adjusted to give good curve fitting. Here, $S_{11}$ has a negative sign because it is compressive.

Experimental stress values and corresponding stretch ratios of each anatomical region were used to perform a non-linear least-square fit of the parameters for three common hyperelastic constitutive models, presented in the next sections. The constitutive models were fitted to the stress-stretch data for strains up to 30%. The fitting was performed using the ***lsqcurvefit.m*** function in MATLAB, and the quality of fit for each model was assessed based on the goodness of the Coefficient of determination $R^2 = \frac{S_t - S_r}{S_t}$, where $S_t =$ the total sum of the squares of the residuals between the data points and the mean and $S_r =$ sum of the squares of the residuals around the regression line. The fitting of hyperelastic models has been comprehensively covered by Ogden et al. (2004), for example.

### 3.2 Fung Strain Energy Function

The Fung isotropic strain energy (Fung, 1967; Fung et al., 1979) is often used for the modelling of soft biological tissues in tension. It depends on the first strain invariant only, as follows,

$$W = \frac{\mu_o}{2b}\left[e^{b(I_1-3)} - 1\right] \tag{6}$$

It yields the following nominal stress component $S_{11}$ along the $x_1$ – axis,

$$S_{11} = \mu_o e^{b(\lambda^2 + 2\lambda^{-1} - 3)}(\lambda - \lambda^{-2}) \tag{7}$$

Here $\mu_o > 0$ (infinitesimal shear modulus) and $b > 0$ (stiffening parameter) are the two constant material parameters to be adjusted in the curve-fitting exercise.

### 3.3 Gent Strain Energy Function

The Gent isotropic strain energy (Gent, 1996) describes rapidly strain-stiffening materials in a very satisfying way. It also depends on the first strain invariant only, as

$$W(I_1) = -\frac{\mu_o}{2} J_m \ln\left(1 - \frac{I_1 - 3}{J_m}\right) \tag{8}$$

It yields the following nominal stress $S_{11}$ along the $x_1$ – axis

$$S_{11} = \frac{\mu_o J_m}{J_m - \lambda^2 - 2\lambda^{-1} + 3}(\lambda - \lambda^{-2}) \tag{9}$$

Here $\mu_o > 0$ (infinitesimal shear modulus) and $J_m > 0$ are two constant material parameters to be optimized in the fitting exercise.

### 3.4 Ogden Strain Energy Function

The Ogden model (Ogden, 1972) has been used in the past to describe the nonlinear mechanical behavior of the brain, as well as of other nonlinear soft tissues (Brittany and Margulies, 2006; Lin et al., 2008; Miller and Chinzei, 2002; Prange and Margulies, 2002; Velardi et al., 2006). Soft biological tissue is often modeled well by the Ogden formulation and most of the mechanical test data available for brain tissue in the literature are fitted with an Ogden hyperelastic function. The one-term Ogden hyperelastic function is given by

$$W = \frac{2\mu_o}{\alpha^2}\left(\lambda_1^\alpha + \lambda_2^\alpha + \lambda_3^\alpha - 3\right) \qquad (10)$$

It yields the following nominal stress $S_{11}$

$$S_{11} = \frac{2\mu_o}{\alpha}\left\{\lambda^{\alpha-1} - \lambda^{-\left(\frac{\alpha}{2}+1\right)}\right\} \qquad (11)$$

Here $\mu_o > 0$ is the infinitesimal shear modulus, and $\alpha$ is a stiffening parameter.

# 4 Results

## 4.1 Stress – Strain Behavior of Brain Tissue

The cylindrical brain samples containing mixed white and gray matter, excised in an anterior – posterior direction from the coronal plane as shown in Fig. 1, were compressed at 30, 60 and 90/s strain rates up to 30% strain. The main objective was to investigate the behavior of tissue under variable loading conditions at rates comparable to impact conditions. Preliminary force - time data obtained at each strain rate was recorded at a sampling rate of 10 kHz. The force (N) was divided by the surface area in the reference configuration to determine the compressive nominal stress (Pa). Fig. 4, shows a typical stress - strain histories curve of brain tissue at a loading velocity of 450 mm/s (90/s). The linear increase in strain with respect to time shows that uniform velocity was achieved during compression. The stress was also determined up to 30% strain: this shows an inherent non linear behavior of brain tissue.

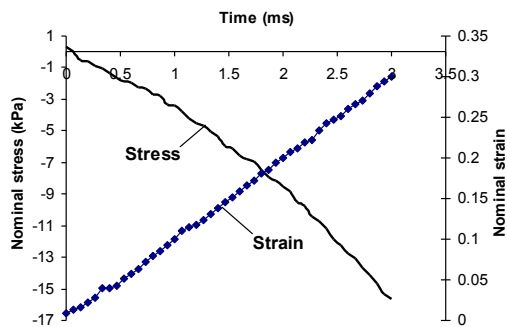

Fig. 4 – Typical stress – strain histories curve at maximum compression velocity of 450 mm/s

## 4.2 Unconfined Compression Tests

Ten tests were performed at each strain rate as shown in Fig. 5, in order to investigate experimental repeatability and behavior of tissue at a particular loading velocity. The tissue stiffness increases with the increase in loading velocity, indicating the strong stress – strain rate dependency of brain tissue. Since the displacement vs. time was also measured simultaneously at a high sampling rate (10 kHz), it was convenient to convert average nominal stress – time data to average nominal stress – strain curve for further analysis. The maximum compressive nominal stress at 30% strain at strain rate of 30, 60 and 90/s are 8.83 ± 1.94 kPa, 12.8 ± 3.10 kPa and 16.0 ± 1.41 kPa (mean ± SD), respectively.

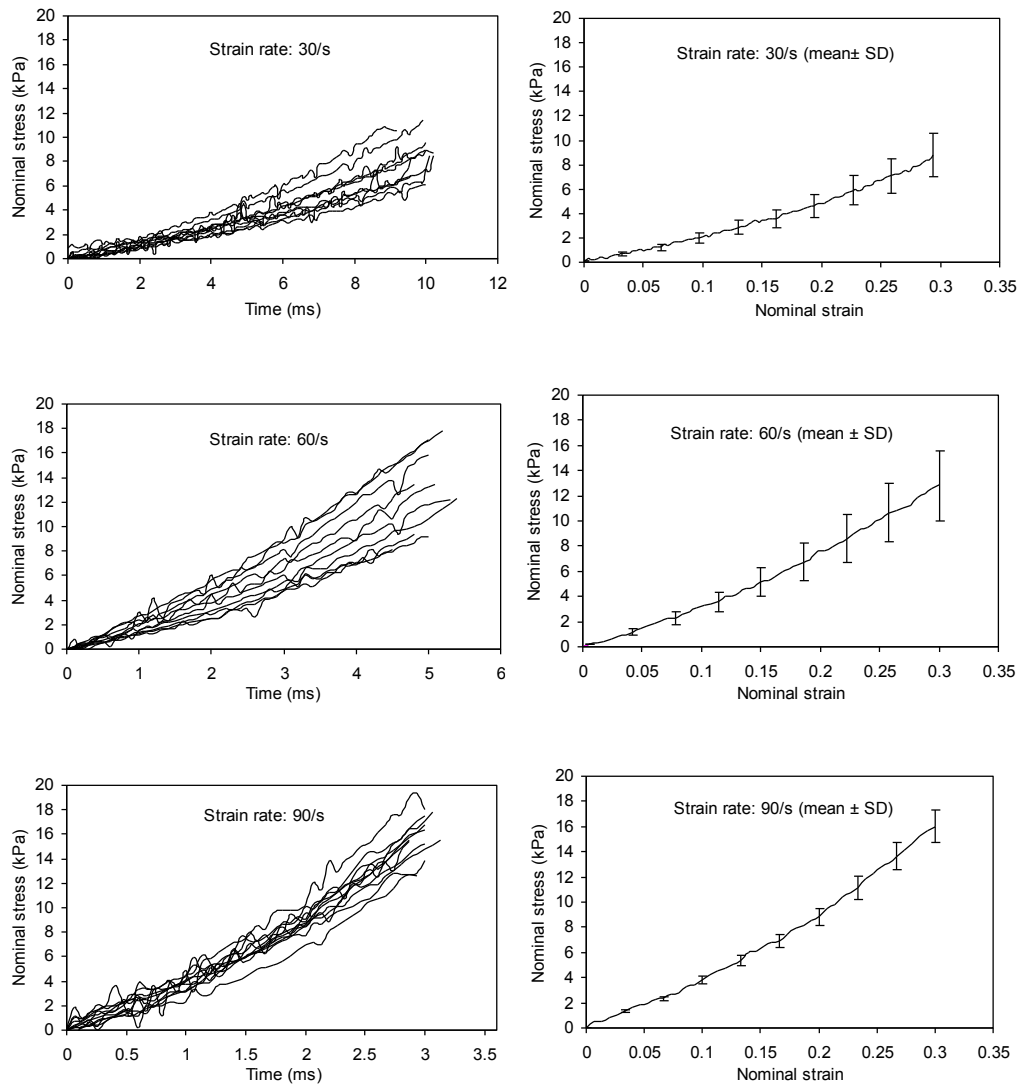

Fig. 5 – Results of ten tests at 30, 60, 90/s strain rates corresponding to compression velocity of 150, 300, and 450 mm/s respectively.

The increase in nominal stress is 32% and 20% from 30 to 60/s and 60 to 90/s strain rates respectively, at 30% strain. The homogeneous expansion of cylindrical brain tissue during compression was also analyzed by a using high speed digital camera (Phantom V5.1 CMOS 10 bit Sensor). The high speed camera images were used to analyze the expansion of specimens in the radial direction during compression. The images were also used to monitor the top and bottom surfaces of each specimen, to ensure that they did not stick to the platens and to confirm that a nearly pure slip condition was achieved. The expansion of cylindrical specimens in the radial direction was approximately homogeneous as shown in Fig. 6. The frequent use of PBS solution on the lower and upper (compression) platens before each test was very effective in reducing friction between the brain tissue and the platens. Similar behavior of the tissue was observed in all experiments. Polytetrafluoroethylene (PTFE) spray lubricant is also a good substitute to reduce friction between the two sliding surfaces. The maximum expansion of the tissue in the middle of the specimen was approximately 1.8 ± 0.1 mm (mean ± SD).

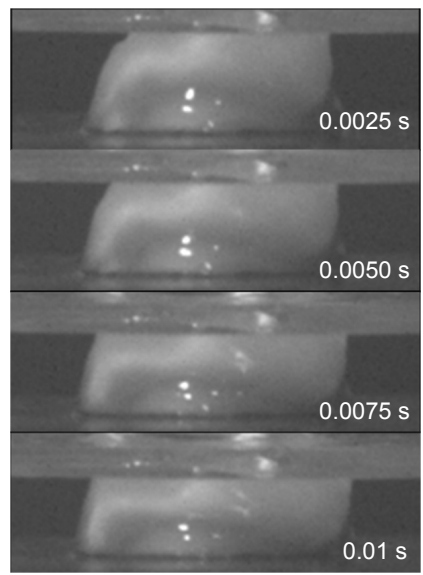

Fig. 6 – Compression stages of the cylindrical brain specimen in terms of time points 0.0025, 0.005, 0.0075 and 0.01s and corresponding stretches at 0.925, 0.85, 0.775 and 0.70, respectively.

### 4.3 Fitting of Constitutive Models to Experimental Data

The average compressive nominal stress – strain curves at each loading rate as shown in Fig. 5, was used for fitting of stress – stretch data to hyperelastic isotropic constitutive models (Fung, Gent and Ogden models). Fitting of each constitutive model to experimental data is shown in Fig. 7: an excellent fit is achieved for all models (coefficient of determination: $0.9988 < R^2 \leq 0.9992$) and the resulting theoretical curves are indistinguishable from each another. The negative sign of the nominal stress indicates compression of brain tissue.

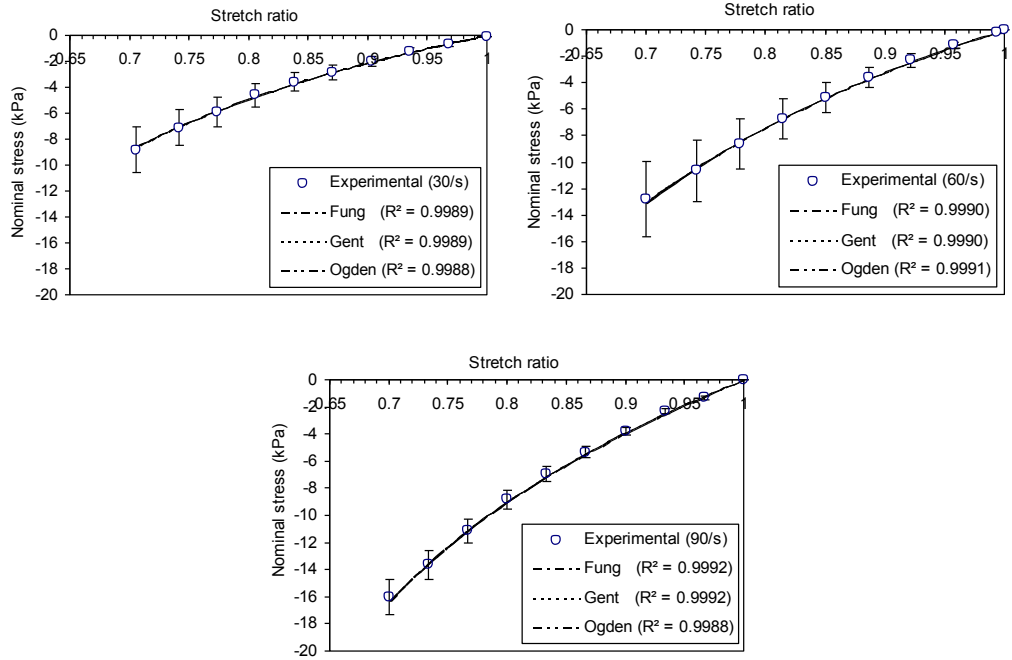

Fig. 7 – Fitting of constitutive models to dynamic strain rate experiments

All of the best fit material parameters ($\mu_o, b, J_m, \alpha$) derived during the curve fitting process for the three constitutive models at each strain rate are summarized in Table 1.

**Table 1** – Material parameters derived after fitting of models to experimental data.

| Strain rate (1/s) | Fung model | | | Gent model | | | Ogden model | | |
|---|---|---|---|---|---|---|---|---|---|
| | $\mu_o$ | $b$ | $R^2$ | $\mu_o$ | $J_m$ | $R^2$ | $\mu_o$ | $\alpha$ | $R^2$ |
| 30 | 6.2753 ± 1.62 | 0.156 ± 0.42 | 0.9989 | 6.2772 ± 1.42 | 6.5568 ± 1.35 | 0.9989 | 6.0568 ± 1.44 | 0.5926 ± 2.5 | 0.9988 |
| 60 | 9.6769 ± 2.63 | 0.014 ± 0.14 | 0.9990 | 9.6773 ± 2.5 | 70.927 ± 9.5 | 0.9990 | 9.4385 ± 2.42 | 1.3551 ± 1.83 | 0.9991 |
| 90 | 11.575 ± 1.64 | 0.147 ± 0.38 | 0.9992 | 11.581 ± 1.28 | 7.0184 ± 1.79 | 0.9992 | 12.642 ± 1.23 | 5.0507 ± 2.6 | 0.9988 |

All $\mu$ are in kPa (mean ± SD) and $\mu > 0$.

It is satisfying to note that the data for the three models coincide in the initial (linear) stage of the testing, and they indicate that the initial shear modulus $\mu_o$ increases with the strain rate. If we consider for instance the Ogden model, we see that the initial shear modulus $\mu_o$ increases from 6.06 to 9.44 kPa (56% increase) from 30 to 60 /s strain rate, with a further increase of 34% from 60 to 90 /s strain rate. Similar values and increase in $\mu_o$ are also observed with the Fung and Gent models. The significant increase in $\mu_o$ with increasing strain rate clearly indicates that a *non-linear viscoelastic* model is required here. The apparent elastic moduli at each strain rate were also calculated from the mean stress – strain curves as shown in Fig. 5, which are summarized in Table 2. The apparent elastic moduli $E_o$, $E_1$ and $E_2$ calculated from the tangent to the stress–strain curve corresponded to the strain ranges of 0 – 0.1, 0.1 – 0.2 and 0.2 – 0.3, respectively.

**Table 2** – Apparent elastic moduli, $E_0$, $E_1$ and $E_2$, of brain tissue at each strain rate (mean ± SD). $p<0.01$ for each modulus value at each strain rate, based on 10 test samples at each strain rate.

| Strain rate (1/s) | $E_0$ (kPa) | $E_1$ (kPa) | $E_2$ (kPa) |
|---|---|---|---|
| 30 | 19.0±0.4 | 28.6±1.3 | 40.5±1.9 |
| 60 | 28.2±0.8 | 48.5±1.9 | 52.0±3.1 |
| 90 | 37.9±0.8 | 56.9±1.2 | 65.2±1.4 |

Hence $E_o = 3\mu_o$ corresponds to the initial Young modulus, while $E_1$ and $E_2$ give a measure of the material stiffness in the intermediate and latter stages of the testing. There is a significant increase in elastic moduli with the increase in strain rate (30 – 90/s), which confirms rate dependency of the tissue. Single factor ANOVA test shows that there is significant difference between the apparent elastic moduli at each strain rate as well as between different strain rates at each elastic moduli ($p < 0.01$), as shown in Table 2. From a phenomenological point of view, we see that the one-term Ogden model is the most apt of the three models at capturing the stiffening of the material with increasing strain rates over the finite range of compression, because it is the only model which has a monotonic variation of its stiffening parameter: hence $\alpha$ increases as the strain rate increases. The stiffening parameter $b$ of the Fung model decreases first and then increases, instead of increasing throughout; and $J_m$ for the Gent model increases first and then decreases instead of decreasing throughout. These deficiencies are a reflection of the limitations of the curve fitting exercise, and also of the poor performance of the Fung and Gent models in the compressive region.

### 4.4 Relaxation Experimentation

The cylindrical specimens were compressed at various strain levels (10% - 50% strain) and held at the same position to record the relaxation force for a short time duration of 500 ms. Since the brain tissue relaxed on the time scale of the ramp, it was necessary to include the ramp loading phase also in each relaxation test. In Fig. 8, the step loading of brain tissue in compression during relaxation at various strain levels is shown.

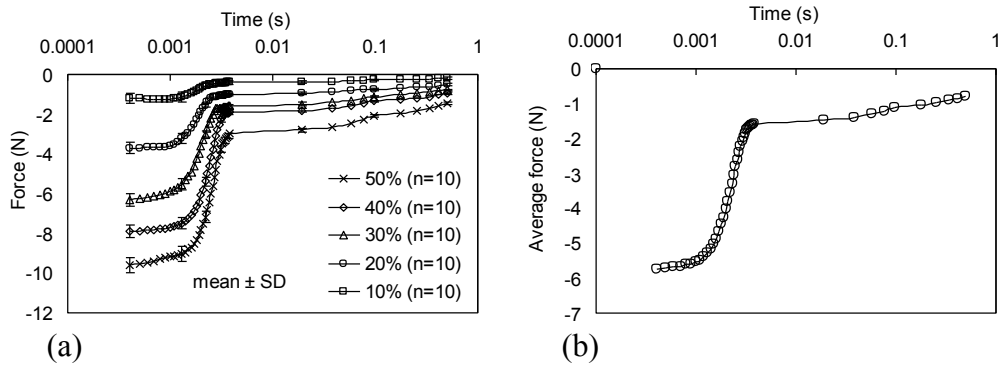

Fig. 8 – Stress relaxation experiments in compression (a) relaxation force (N) at various strain magnitudes (b) average relaxation force.

The compressive force decreased by approximately 70% (average of 10% to 50% strain) within 4.0 ms of the relaxation time, then it continuously decreased gradually up to 500 ms. The dramatic decrease in force up to 4 ms reveals the highly viscoelastic nature of brain tissue.

### 4.5 Ogden-based Hyper-viscoelastic model

The elastic and viscoelastic behavior of brain tissue can be characterized using an Ogden based non-linear viscoelastic model. This is a single integral, finite strain, isotropic model. The relaxation response is based on a Prony series expansion and the strain energy function is developed in the form of a convolution integral, used earlier by various research groups (Miller and Chinzei, 2002; Prange et al., 1999; Prange and Margulies, 2002).

$$W = \frac{2}{\alpha^2} \int_0^t \left[ \mu(t-\tau) \frac{d}{d\tau} (\lambda_1^\alpha + \lambda_2^\alpha + \lambda_3^\alpha - 3) \right] d\tau \qquad (12)$$

Hence, the relaxation of the time-dependent shear modulus $\mu(t)$ to describe the viscous response of the tissue is,

$$\mu(t) = \mu_0 \left[ 1 - \sum_{k=1}^{n} g_k (1 - e^{-t/\tau_k}) \right] \tag{13}$$

where $\mu_0$ is the initial shear modulus in the undeformed state, $\tau_k$ are the characteristic relaxation times, and $g_k$ are the relaxation coefficients, which can be determined from the experimental data. The average experimental relaxation data (shown in Fig. 8 (b)) and Ogden hyperelastic parameters (given in Table 1) were used to analyze the behavior of brain tissue from the response curves generated in ABAQUS 6.9/Explicit as discussed in Analysis User's Manuel Section 18.7.1. The average estimated prony parameters were $g_k$ = 0.8220 and $\tau_k$ = 0.00298 s with goodness of fit 7.15 (percent of root mean square error).

## 5 Finite Element Analysis

### 5.1 Numerical and Experimental Results

Numerical simulations were performed by applying various boundary conditions using ABAQUS 6.9/ Explicit in order to mimic experimental conditions. The mass density $1040\,kg/m^3$ and material parameters listed in Table 1 for the Ogden strain energy function were used for numerical simulations. 9710 hexagonal C3D8R elements (8-node linear brick, reduced integration with relax stiffness hourglass control) were used for the brain part. During the simulations, the top surface of the cylindrical specimen was compressed in order to achieve 30% strain at various loading velocities (150, 300 and 450 mm/s). Reaction forces (N) on the bottom surface of the cylindrical specimen were added after the simulation and divided by the cross sectional area of the specimen in the undeformed configuration to calculate the engineering stresses (kPa). An excellent agreement of the average experimental stresses with the numerical stresses (kPa) was achieved as shown in Fig. 9 (a). A statistical analysis based on one-way ANOVA showed that there was insignificant difference (p = 0.9985, p = 0.9199, p = 0.9389) between the experimental and numerical stresses at 30, 60 and 90/s strain rates, respectively. Figure 9 (b) shows both deformed and undeformed states of the cylindrical brain specimen and the reaction forces (N) in the z-direction at 30% strain; at a strain rate of 30/s. During the numerical simulations, it was important to analyze artificial strain energy or "hourglass stiffness" which is used to control hourglass deformation. It was observed that accumulated artificial strain energy (ALLAE) for the whole model as percentage of the total strain energy was 0.0036% for the simulations at 30/s strain rate. The magnitudes of various energies (J) of the numerical model ($\mu_0$ = 6056.8 Pa, $\alpha$ = 0.5926 at 30/s strain rate from Table 1) are also shown on a logarithmic scale in Fig. 9 (c). Similarly, the ALLAE as percentage of the total strain energy was 0.0027% and 0.0025% for the simulations performed at 60 and 90/s strain rates, respectively. The significant low percentage of artificial strain energy ($\leq$ 0.0036%) observed during the simulations at all strain rates (30, 60 and 90/s) indicates that hourglassing is not a problem.

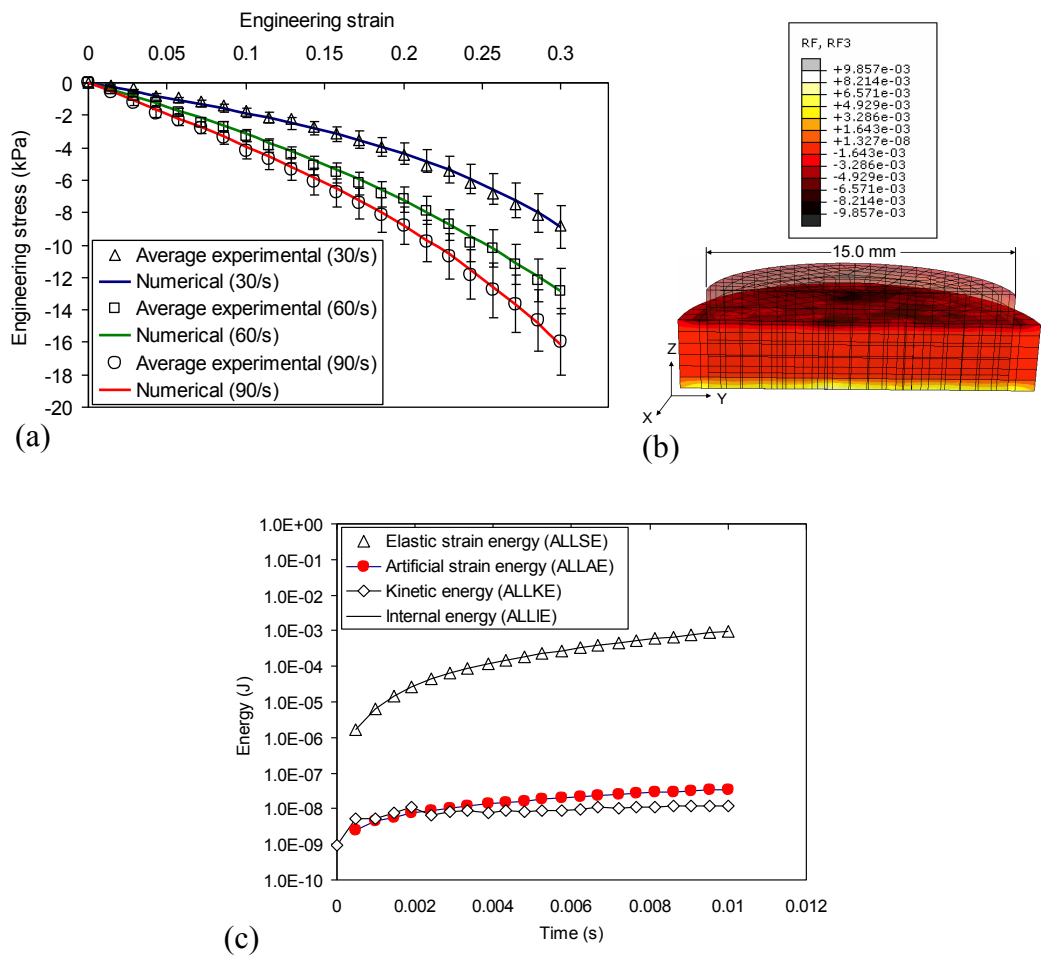

Fig. 9 – Numerical simulations at various strain rates and energies involved (a) agreement of experimental and numerical engineering stresses in unconfined compression (error bars indicate SD) (b) Reaction forces (N) observed at 30% compression using Ogden material parameters ($\mu_0$ = 6056.8 Pa, $\alpha$ = 0.5926 from Table 1), showing deformed and undeformed state of the brain specimen (c) various strain energies (J) observed in the brain specimen at a strain rate of 30/s.

## 5.2 Determination of Friction Coefficient

A combined experimental and computational approach was necessary in order to estimate the amount of friction generated during unconfined compression of brain tissue. A separate set of lubricated and bonded experiments under the same test conditions were performed according to procedure mentioned in Section 2.6. The engineering stresses determined experimentally at a loading velocity of 150 mm/s (30/s) under lubricated and bonded conditions were 8.83 ± 1.80 kPa and 18.9 ± 4.53 kPa (mean ± SD), respectively, at 30% strain as shown in Fig.10 (a). Similarly, the experimental engineering stresses at a loading velocity of 450 mm/s (90/s) under lubricated and bonded conditions were 15.86 ± 3.25 kPa and 31.57 ± 7.58 kPa (mean ± SD), respectively, at 30% strain as shown in Fig. 10 (b). The numerical simulations were performed in ABAQUS 6.9/ Explicit after creating the geometry of a cylindrical brain specimen, top and lower platens. A *rigid body constraint* was applied to the platens and *kinematic contact method* was applied between the brain specimen and platen surfaces. The platen surface was selected as master surface, and the brain surface as slave. The *penalty* option was used to estimate the stresses at various $\mu$ values. The time period for each simulation was adjusted to achieve the required amount of compression (in our case, 30% compression) for the brain specimen, which was directly linked to the velocity of the top platen during the experiments. Several iterations were performed by assuming $\mu = 1$ and arbitrary Ogden material parameters. After few simulations, an excellent agreement between bonded (experimental) and numerical stresses was achieved at $\mu_0 = 5600$ Pa, $\alpha = 2.7$ as shown in Fig. 10 (c). Thereafter, material parameters were kept constant and only $\mu$ was varied until numerical stresses were in good agreement at $\mu = 0.1 \pm 0.03$ against stresses under lubricated conditions 8.83 ± 1.80 kPa (mean ± SD) at 30% strain and 30/s strain rate as shown in Fig. 10 (c). A similar procedure was also adopted to estimate $\mu$ at 90/s strain rate. The bonded (experimental) and numerical stresses were in good agreement at $\mu = 1$ using material parameters as $\mu_0 = 9000$ Pa, $\alpha = 3.0$ as shown in Fig. 10 (d). While keeping material parameters constant and after several iterations, stresses under lubricated conditions, 15.86 ± 3.25 kPa at 30% strain, were in good agreement with numerical stresses at $\mu = 0.15 \pm 0.07$ (mean ± SD) as shown in Fig. 10 (d).

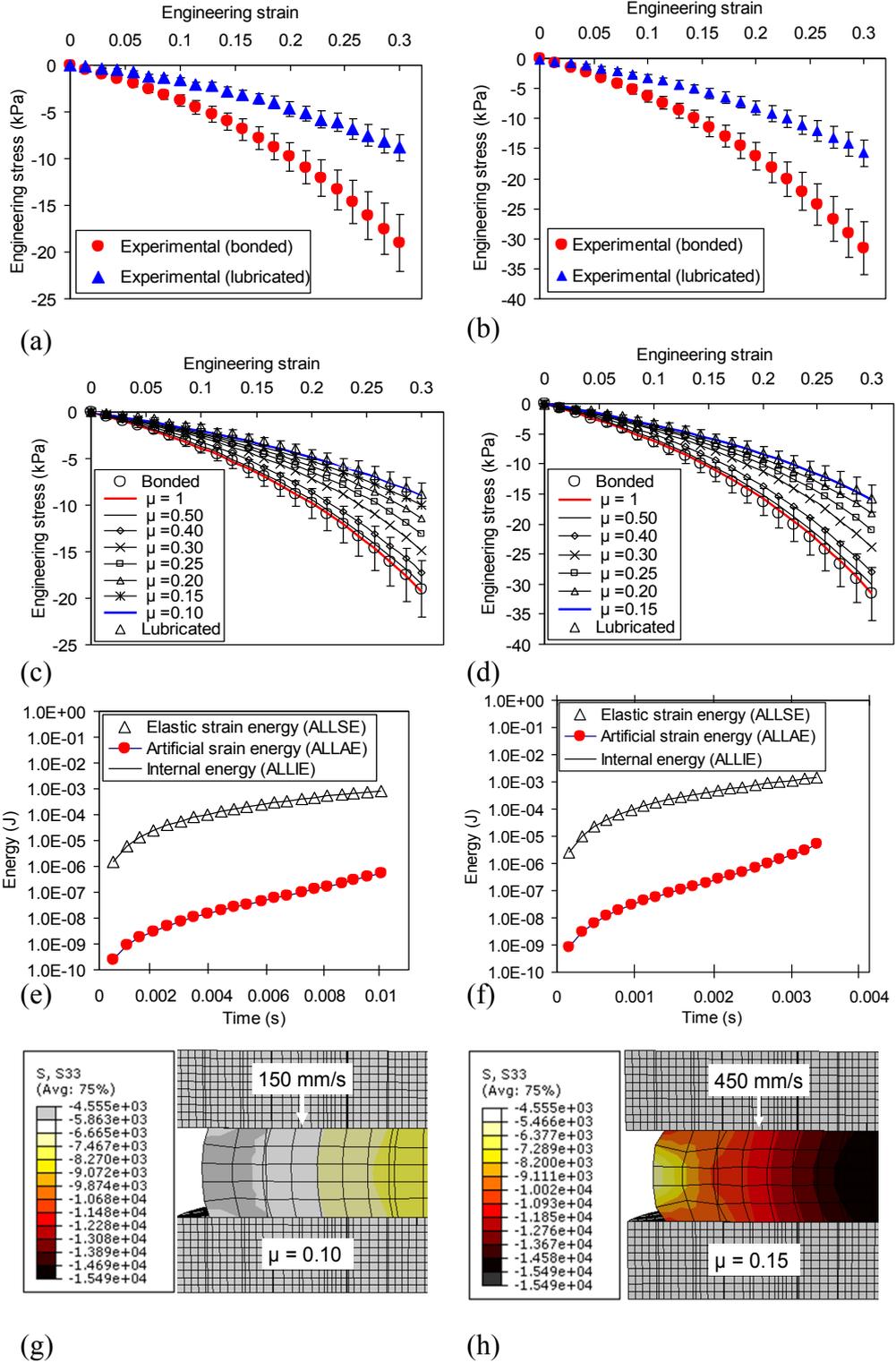

Fig. 10 – Lubricated and bonded stresses up to 30% strain (a) at 30/s strain rate (b) at 90/s strain rate (c) simulations: $\mu_0$ = 5600 Pa, $\alpha$ = 2.7 at 30/s strain rate at ( $\mu$ = 1 and $\mu$ = 0.1 ± 0.03) (d) simulations: $\mu_0$ = 9000 Pa, $\alpha$ = 3.0 at 90/s strain rate ( $\mu$ = 1 and $\mu$ = 0.15 ± 0.07) (e) Energies at 30/s strain rate (f) Energies at 90/s strain rate (g) stress contours at μ = 0.10 at 30/s strain rate (h) stress contours at μ = 0.15 at 90/s strain rate.

The magnitudes of various energies of the whole numerical model were also determined to analyze hourglass stiffness affects. The artificial strain energy (ALLAE) as percentage of the total strain energy was 0.067% and 0.355% at a strain rate of 30 and 90/s respectively (Fig. 10 (e) and (f)). The significant low percentage of artificial strain energy (≤ 0.355%) indicates that hourglassing is not a problem during simulations for the determination of friction coefficient. Typical stress contours and inhomogeneous deformation of the brain specimen at mean values of friction coefficient ($\mu = 0.10$ and $\mu = 0.15$) are shown in Fig. 10 (g) and (h). The estimated range of $\mu = 0.13 – 0.22$ (maximum) through combined experimental and computational approach indicates that pure slip conditions cannot be achieved, even under fully lubricated test conditions. One-way ANOVA test was carried out for the statistical comparison of experimental and the numerical engineering stresses using data shown in Fig. 10 (a) – (d). There was insignificant difference (p = 0.7627) between the stresses under lubricated conditions and at $\mu = 0.1$ as shown in Fig. 11 (a), similarly p = 0.9798 for the stresses under bonded condition and at $\mu =1$, however p = 0.00152 for the bonded and lubricated engineering stresses indicating significant difference as shown in Fig. 11 (a). Similar statistical differences were also observed in the case of 90/s strain rate as shown in Fig. 11 (b).

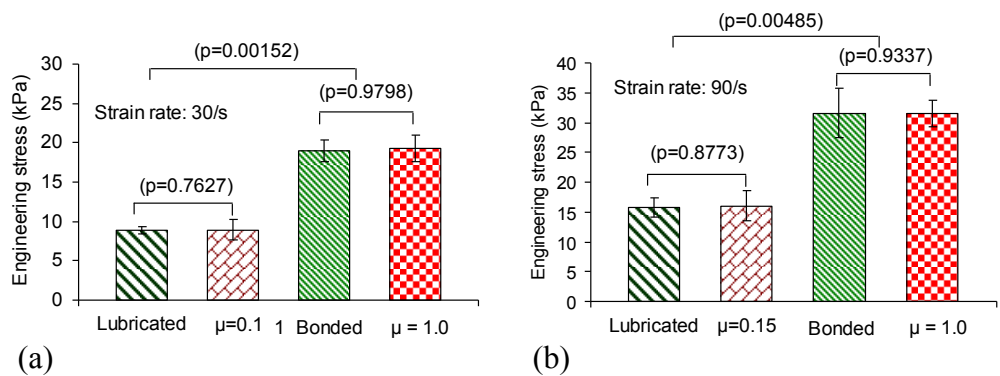

Fig. 11 – Comparative analysis of experimental (bonded, lubricated) and numerical engineering stresses (a) comparison at 30/s strain rate (b) comparison at 90/s strain rate

# 6    Discussion

In the present study, the mechanical properties of porcine brain tissue have been determined during unconfined compression of brain tissue at intermediate strain rates (30 – 90/s). The compressive nominal stress at 30% strain was 8.83 ± 1.94 kPa, 12.8 ± 3.10 kPa and 16.0 ± 1.41 kPa (mean ± SD) at strain rates of 30, 60 and 90/s respectively, which shows the high rate dependency of brain tissue. The average experimental data at each strain rate was used to determine material parameters by using one-term Ogden, Fung and Gent strain energy functions. For the one-term Ogden model, the initial shear modulus $\mu_o$ is 6.06 ± 1.44, 9.44 ± 2.427 and 12.64 ± 1.227 kPa (mean ± SD) at strain rates of 30, 60 and 90/s respectively; see Table 1. Moreover, relaxation tests were also performed from 10% - 50% strain with an average rise time of 10 ms for further hyperviscoelastic analysis of brain tissue. Excellent agreement between the experimental and numerical engineering stresses (Fig. 9 (a)) shows that the Ogden strain energy function is fully able to characterize the behavior of brain tissue in compression. Artificial strain energy as a percent of the total strain energy was observed to be insignificant (≤ 0.355%) during the numerical simulations; therefore reported numerical results can be used with confidence.

The compressive nominal stress at 30% strain and apparent elastic moduli (strain range: 0 – 0.2) are considered for comparison purposes and are summarized in Table 3.

Table 3 – Variation of parameters with the change in strain rates

| Strain rate | Compressive nominal stress at 30% strain (kPa) | Apparent elastic moduli, $E$ strain (0 – 0.2) (kPa) | Temperature conditions during test |
|---|---|---|---|
| 40/s Estes and McElhaney (1970) | ~ 26.4* | ~ 41.7** | 37 °C |
| 50/s (Tamura et al., 2007) | ~ 11.0 | 23.8 ± 10.5 | 20 °C |
| 60/s (present study) | 12.8 ± 3.10 | 38.5 ± 2.0 | 22 °C |

*Compressive true stress **Tangent shear modulus at 10% strain (Mendis et al., 1995)

With the increase in strain rate, the compressive nominal stress and apparent elastic moduli also increase, as brain tissue is strain rate dependent. There is a 16% increase in compressive nominal stress and 61.76% increase in apparent elastic moduli (if mean values only are considered), because of the increase in strain rate from 50 to 60/s. This proportional increase is expected because of the increase in strain rates. However, the results of Estes and McElhaney (1970) are much higher, even at a lower strain rate of 40/s. The reasons for these high stress values are still not known, although similar variations in results were also noticed by Tamura et al. (2007).

A combined experimental - computational approach was adopted to determine values of the friction coefficient. Before numerical simulations, it was essential to perform both lubricated and bonded unconfined compression tests under the same controlled environment. The values of $\mu$ ranged from 0.07 to 0.22 and the average value of $\mu$ was 0.13 ± 0.05 (mean ± SD). The values of $\mu$ determined in this study are fundamentally dependent on various conditions (Coulomb friction model available in ABAQUS/6.9 Explicit, high precision in experimental data both under lubricated and bonded conditions, careful selection of boundary conditions and types of surface interactions for the numerical simulations). To the best of authors' knowledge, there is no study available to compare the values of friction coefficient estimated in this study. A study conducted by Wu et al. (2004) found that the stress of soft tissue specimens (pigskin, pig brain, and human calcaneal fat) obtained from the specimen/platen friction can be overestimated by 10 – 50% with the frictionless specimen/platen contact, even in well-lubricated test conditions. Moreover, a study conducted by Zhang and Mak (1999) reported frictional properties of skin and found values of friction coefficient as 0.46 ± 0.15 (mean ± SD).

To analyze the effects of stress wave propagation on experimental results, unconfined compression tests were performed on porcine brain tissue at a maximum strain rate of 90/s. Cylindrical specimens of nominal dimensions of 15.0 mm diameter and variable thickness of 3.0, 4.0 and 5.0 mm from mixed white and gray matter were prepared using the procedure mentioned in Section 2.1. The compression platen velocity was 270, 360, 450 mm/s against maximum compressive displacement of 0.9, 1.2, 1.5 mm respectively for 3.0, 4.0 and 5.0 mm thick specimens in order to achieve a constant strain rate of 90/s. The measured engineering stresses (kPa) of each specimen were compared as shown in Fig. 12. It was observed that the average peak stress was 21.51 ± 2.95 kPa, 21.0 ± 1.97 kPa, 20.57 ± 3.09 kPa (mean ± SD). There was a slight decrease in peak stress values (approximately 2%) with the increase in specimen thickness from 3.0 to 4.0 mm and from 4.0 to 5.0 mm as shown in Fig. 12. Thus, it was observed that the experimental results were not overestimated due to stress wave propagation effects. The single factor ANOVA test was carried out to determine variations between different groups of specimen thickness and the degree of variation with in each group. There was no significant difference in engineering stresses (p = 0.5802) between different specimen thicknesses (3.0, 4.0 and 5.0 mm) as shown in Fig. 12.

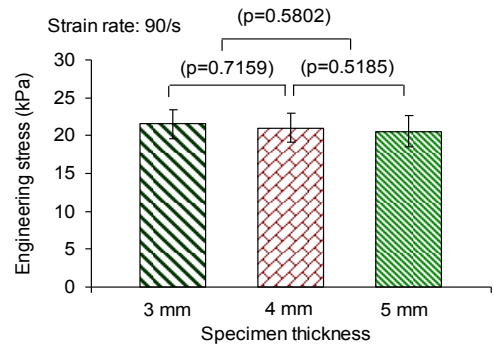

Fig. 12– Comparison of engineering stresses at variable specimen thickness of brain tissue to investigate effects of stress wave propagation at a strain rate of 90/s

In this study, we have conducted unconfined compression and relaxation tests at a room temperature of ~22 °C. Miller and Chinzei (1997, 2002) also performed such tests at the same temperature while, Shen et al. (2006) tested at 37 °C to simulate in vivo conditions. They also performed oscillatory tests at 30 °C, 20 °C, and 10 °C to extend the data over a wide frequency range by using Time Temperature – Superposition (TTS) principle. Another important study was conducted by Hrapko et al. (2008) to analyze the difference between room temperature (approximately 23 °C) and body temperature (approximately 37 °C) conditions and to scale results obtained at these different conditions. The measured results were found to be clearly temperature dependent and the dynamic modulus $G^*$, at 23 °C, was approximately 35% higher than at 37 °C. Stiffening of the samples occurred with decreasing temperature. Zhang et al. (2011) conducted tests on porcine brain tissue at high – strain rates specifically to investigate stress – strain behavior at ice cold temperature and at 37 °C. The estimated stresses at 37 °C were 60 – 70% higher than at ice cold temperature, showing a stiffer response of brain tissue at higher temperature (37°C). These findings are in direct contradiction to the research conducted by Hrapko et al. (2008), which showed stiffer response of brain tissue at the lower temperature (23 °C). Based on these contradictory findings, there is a crucial need to further investigate the effects of temperature on brain tissue.

The limitation of this study is that *in vitro* tests were performed on Porcine brain tissue. The variation in results may be due to the difference between *in vivo* and *in vitro* properties of the tissue. Gefen and Margulies (2004) carried out comparison between *in vivo* and *in vitro* mechanical behavior of brain tissue and found that the postmortem time for testing was considered as the dominant cause for the large variation in results, whereas pressurized vasculature (during *in vivo* tests), loss of perfusion pressure (during *in vitro* tests) and inter-species variability have very little effect on the experimental results. On the basis of interesting comparisons between the *in vitro* and *in vivo* tests, the shear modulus estimated from the *in vitro* rheometric data (Nicolle et al., 2004) was approximately within the same range of *in vivo* MRE experimental results (McCracken et al., 2005). However, there is still a need to do further research to confirm the interchangeability of *in vitro* and *in vivo* results

Another limitation of this study is the estimation of material parameters from the strain energy functions, based on the average mechanical properties (mixed white and gray matter) of the brain tissue; however, these results are still useful in the approximate behaviour of brain tissue. Moreover, the average mechanical properties were also determined by Miller and Chinzei (1997, 2002). In previous studies, it was observed that the anatomical origin or location as well as direction of excision of samples (superior – inferior and medial – lateral direction) had no significant effect on the results (Tamura et al., 2007) and similar observations were also reported by Donnelly and Medige (1997). Tensile tests were also conducted using mixed white and gray matter up to a maximum strain rate of 25/s (Tamura et al., 2008). Based on the research conducted by Prange and Margulies (2002), by using samples of size 55 mm x 10 mm and 1 mm thick, the gray matter showed no difference between the two orthogonal directions whereas the white matter showed significantly different behaviour. Similarly, in the case of directional properties across regions, comparisons were made between the corona radiata and corpus callosum. The corona radiata was significantly stiffer than the corpus callosum, and the white matter behavior was more anisotropic than gray matter. In our future research, we intend to characterize the mechanical behaviour of white and gray matter as well as regional and directional properties at intermediate strain rates (1, 10, 20, 30/s) using indentation methods.

# 7. Conclusions

The following results can be concluded from this study:

1 – The estimated compressive nominal stress at 30% strain is 8.83 ± 1.94 kPa, 12.8 ± 3.10 kPa and 16.0 ± 1.41 kPa (mean ± SD) at strain rates of 30, 60 and 90/s, respectively.

2 – One-term Fung, Gent and Ogden models provide excellent fitting to experimental data up to 30% strain (coefficient of determination: $0.9988 < R^2 \leq 0.9992$).

3 – The Ogden model is readily available in ABAQUS/6.9 and can be used efficiently for finite element simulations. Excellent agreement between the experimental and numerical engineering stresses indicates that Ogden strain energy function is fully capable to characterize the behavior of brain tissue in compression.

4 – Combined experimental (bonded and lubricated tests) and computational approach can be adopted to determine values of friction coefficient. The estimated values of $\mu$ at the brain specimen/platen interface are $\mu = 0.1 \pm 0.03$ and $\mu = 0.15 \pm 0.07$ (mean ± SD) at 30 and 90/s strain rates, respectively.

5 – The test apparatus developed for the unconfined compression of porcine brain tissue can be used with confidence to extract data at a uniform velocity at strain rates of 30, 60 and 90/s.

6 – This study provides a set of constitutive data at intermediate strain rates, based on which rate-dependent material models may be developed.

**Acknowledgements**    The authors thank Dr Manuel Forero for his guidance with ABAQUS simulations, Aisling Ní Annaidh for her assistance with curve fitting by using MATLAB, and John Gahan, Tony Dennis and Pat McNally for their assistance in machining components and developing electronic circuits for the experimental setup. This work was supported for the first author by a Postgraduate Research Scholarship awarded by the Irish Research Council for Science, Engineering and Technology (IRCSET), Ireland.